\documentclass[sn-mathphys,Numbered]{sn-jnl}

\usepackage{graphicx}%
\usepackage{multirow}%
\usepackage{amsmath,amssymb,amsfonts}%
\usepackage{amsthm}%
\usepackage{mathrsfs}%
\usepackage[title]{appendix}%
\usepackage{xcolor}%
\usepackage{textcomp}%
\usepackage{manyfoot}%
\usepackage{booktabs}%
\usepackage{algorithm}%
\usepackage{algorithmicx}%
\usepackage{algpseudocode}%
\usepackage{listings}%

\theoremstyle{thmstyleone}%
\theoremstyle{thmstyletwo}%

\theoremstyle{thmstylethree}%

\begin{document}

\title[Kinetic inductance traveling wave amplifier designs for practical microwave readout applications]{Kinetic inductance traveling wave amplifier designs for practical microwave readout applications}


\author*[1,2,3,4,5]{\fnm{A.} \sur{Giachero}}\email{andrea.giachero@colorado.edu}

\author[1]{\fnm{M.} \sur{Vissers}}
\author[1]{\fnm{J.} \sur{Wheeler}}
\author[1,2]{\fnm{L.} \sur{Howe}}
\author[1,2]{\fnm{J.} \sur{Gao}}
\author[1]{\fnm{J.} \sur{Austermann}}
\author[1]{\fnm{J.} \sur{Hubmayr}}
\author[3,4,5]{\fnm{A.} \sur{Nucciotti}}
\author[1,2]{\fnm{J.} \sur{Ullom}}

\affil[1]{\orgname{National Institute of Standards and Technology}, 
          \orgaddress{\city{Boulder}, 
                      \postcode{80305}, 
                      \state{Colorado}, 
                      \country{USA}
                    }
            }

\affil[2]{\orgdiv{Department of Physics},
          \orgname{University of Colorado}, 
          \orgaddress{\city{Boulder}, 
                      \postcode{80309}, 
                      \state{Colorado}, 
                      \country{USA}
                    }
            }

\affil[3]{\orgdiv{Department of Physics},
          \orgname{University of Milano Bicocca}, 
          \orgaddress{\city{Milan}, 
                      \postcode{I-20126}, 
                      \country{Italy}
                    }
            }

\affil[4]{\orgname{INFN - Milano Bicocca}, 
          \orgaddress{\city{Milan}, 
                      \postcode{I-20126}, 
                      \country{Italy}
                    }
            }

\affil[5]{\orgname{Bicocca Quantum Technologies (BiQuTe) Centre}, 
          \orgaddress{\city{Milan}, 
                      \postcode{I-20126}, 
                      \country{Italy}
                    }
            }            


\abstract{
A Kinetic Inductance Traveling Wave amplifier (KIT) utilizes the nonlinear kinetic inductance of superconducting films, particularly Niobium Titanium Nitride (NbTiN), for parametric amplification. These amplifiers achieve remarkable performance in terms of gain, bandwidth, compression power, and frequently approach the quantum limit for noise. However, most KIT demonstrations have been isolated from practical device readout systems.  Using a KIT as the first amplifier in the readout chain of an unoptimized microwave SQUID multiplexer coupled to a transition-edge sensor microcalorimeter we see an initial improvement in the flux noise~\cite{Malnou2023}. One challenge in KIT integration is the considerable microwave pump power required to drive the non-linearity. To address this, we have initiated efforts to reduce the pump power by using thinner NbTiN films and an inverted microstrip transmission line design. In this article, we present the new transmission line design, fabrication procedure, and initial device characterization -- including gain and added noise. These devices exhibit over 10\,dB of gain with a 3~dB bandwidth of approximately 5.5--7.25\,GHz, a maximum practical gain of 12\,dB and typical gain ripple under 4~dB peak-to-peak. We observe an appreciable impedance mismatch in the NbTiN transmission line, which is likely the source of the majority of the gain ripple. Finally we perform an initial noise characterization and demonstrate system-added noise of three quanta or less over nearly the entire 3\,dB bandwidth.
}




\maketitle

\section{Introduction}
Fundamental physics experiments operating at microwave frequencies require ultra-sensitive readout schemes that greatly benefit from a amplification chain with the lowest possible noise, high gain, and large bandwidth. These requirements are essential for the readout of large arrays of detectors without compromising the information conveyed by the signal. Although semiconductor amplifiers, such as high-electron-mobility transistor (HEMT) amplifiers, provide large gain and bandwidths~\cite{Tracy2018}, their noise is often $\sim 10–40$ times  the Standard Quantum Limit (SQL)~\cite{Caves1982} at microwave frequencies. Josephson parametric amplifiers~\cite{Roy2016,Castellanos2007} (JPAs) offer noise performance at or below the SQL. However, their bandwidth is limited to a few hundred megahertz~\cite{White2023} -- significantly limiting the number of devices they can practically read out. 

An emerging technology for achieving these requirements is that of the broadband TWPA, such as the Josephson TWPA (JTWPA)~\cite{Macklin2015} or the kinetic inductance TWPA (KITWPA)~\cite{HoEom2012}, or KIT. These devices reach noise levels near the SQL with high gain over a larger bandwidth 
(a few GHz).
A TWPA amplifier generally consists of a long transmission line designed to exploit a specific nonlinearity in a superconducting circuit. A large pump tone modulates these nonlinear elements, coupling the pump ($f_p$) to a signal ($f_s$) and idler ($f_i$) tone via frequency conversion. In the Four-Wave-Mixing (4WM) case $2 f_p = f_s + f_i$, while for three-wave-mixing (3WM) $f_p = f_s + f_i$, i.e. abundant pump photons are exchanged for signal (and idler) photons, resulting exponential signal gain as photons interact with the modulated nonlinearity. In the case of JTWPA, the nonlinearity used is the Josephson inductance\cite{OBrien2014}, while a KIT uses the superconducting nonlinear kinetic inductance~\cite{Erickson2017}. Compared with JTWPAs, KITs are simple to fabricate and require only few lithography and etching steps, without overlapping structures. Additionally, KIT amplifiers based on Niobium Titanium Nitride (NbTiN) thin films provide a high dynamic range (-$60$\,dBm), gain (20\,dB), and operate near the SQL~\cite{Malnou2021}. KITs have been successfully used to read out superconducting qubits \cite{Ranzani2018}, Microwave Kinetic Inductance Detectors (MKIDs)~\cite{Zobrist2019}, Transition Edge Sensor (TESs)~\cite{Malnou2023} and also showed impressive performance operating at 4\,K~\cite{Malnou2022}.

 However, contemporary KITs require a strong pump, typically around -30\,dBm at the KIT input~\cite{Malnou2021}. Isolating this power from the device under test is challenging, potentially resulting in excess noise and a degradation of readout channel performance. These devices are based on 20-30\,nm NbTiN films with a kinetic inductance, $L_k$, in the range of 7-10\,pH/sq~\cite{Malnou2021,Shu2021}. Reduction of the pump power may be realized by using a thinner superconducting film with a higher kinetic inductance~\cite{Giachero2023}. In this study, we present preliminary results on KITs designed with a target $L_k$ of 35\,pH/sq adapted from the previous coplanar waveguide geometry~\cite{Malnou2021} to an inverted microstrip~\cite{Shu2021}.

\begin{figure}[!t] 
 \begin{center}
    \includegraphics[clip=true,width=0.9\textwidth]{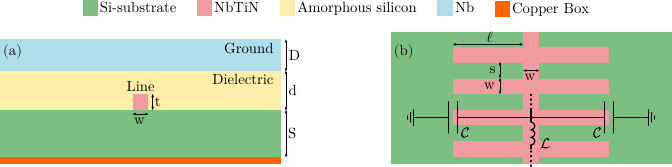}
  \end{center}
 \caption{\label{fig:TRL_lines} Cross-sectional side (a) and top side (b) views of a stub-loaded inverted micro-strip line. Relative dimensions are not to scale.}
\end{figure}

\section{Transmission line design}
The kinetic inductance of a superconducting transmission line under dc current bias $I_{dc}$ is  $L_k(I)=L_d\left(1+{I^2}/{I_*^2}\right)$, with $L_d=L_0\left(1+I_{dc}^2/{I_*^2}\right)$~\cite{Malnou2021}. Here $I_*$ is an intrinsic material parameter that controls the scale of the kinetic inductance nonlinearity~\cite{Zmuidzinas2012}. $I$ ($I_{dc}$) is the rf (dc) current, $L_0$ is the NbTiN kinetic inductance at zero dc current, $L_d$ is the line inductance under nonzero dc bias and zero rf current ($I=0$). $\varepsilon=2I_{dc}/(I_*^2+I_{dc}^2)$ describes the 3WM process, while and $\xi=1/(I_*^2+I_{dc}^2)$ describes 4WM.

The scaling current $I_*$ is directly proportional to the cross-sectional area of the film ($A = t \cdot w$, see Fig.~\ref{fig:TRL_lines}), thus decreasing $t$ reduces $I_*$. Moreover, reducing the film thickness $t$ results in an increase in $L_0$, so the required pump power (dc bias) to achieve a given $L_k$ decreases quadratically (linearly) with $t$. Thicknesses of 5 and 10\,nm yield kinetic inductance values of approximately 100\,pH/sq and 30\,pH/sq, respectively -- along with scaling currents of about 0.6\,mA and 3\,mA, respectively\cite{Giachero2023}. These values should be compared with the previous KIT device: $L_k=10\,$pH/sq and $I_*=7$\,mA~\cite{Malnou2021,Malnou2022}. 

Using an Inverted Microstrip (IMS) geometry, we recently implemented a $t=10$\,nm thick NbTiN film to realize $L_k$ of around 35\,pH/sq. The \mbox{$\alpha$-Si} thickness is $d=100$\,nm. The sky (ground) plane is made of a thick Nb layer ($t=100$\,nm, $L_k$ negligible) deposited on top of the \mbox{$\alpha$-Si}. The microstrip center line and finger widths are $w=1\,\mu$m while the spacing between adjacent fingers is $s=1\,\mu$m, resulting in a elementary cell length of $w+s=2\,\mu$m (Fig.~\ref{fig:TRL_lines}). The line impedance is determined by the inductance (capacitance) per unit length $\mathcal{L}$ ($\mathcal{C}$) via $Z_0 = \sqrt{\mathcal{L} / \mathcal{C}}$. With these dimensions, electromagnetic simulations provided a finger length of $18\,\mu$m ($6.5\,\mu$m) for matching $Z_0=50\,\mathrm{\Omega}$ ($Z_0=80\,\mathrm{\Omega}$). The full KIT line is made from a string of 1200 \textit{super-cells}, each composed of 30 \textit{unloaded-cells} and 6 \textit{loaded-cells}. This gives a stop-band around 10.5--11.00\,GHz and, since the gain profile is centered near $f_p/2$ and the pump is applied just above the stop-band, this centers the gain near 6\,GHz. The total transmission line length is 8.64\,cm. 

Beyond shortening the total line length (and thus overall device size) via increasing the kinetic inductance and, as consequence, the inductance per unit length and the  stub-to-ground capacitance per unit length, the IMS approach also benefits from significantly higher fabrication yield relative to previous devices implementing a Coplanar Waveguide (CPW) transmission line. Elimination of the potential for a center-ground short (lithographic) failure common in the CPW geometry resulted in a 100\% device yield with the new IMS devices. On the other hand, the presence of a non-vacuum dielectric could potentially increase loss and noise, including those sourced from two-level system (TLS) interactions. 
 
\section{Device Fabrication}\label{sec:fap}

The devices are fabricated on a 76.2\,mm, high-resistivity, intrinsically-doped, float-zone Si wafer. Immediately before loading into the vacuum chamber the native oxide is stripped using HF~\cite{wisbey2010-HF}. As for previous designs, the NbTiN is reactively co-sputtered from Ti and Nb targets in an Ar:N$_2$ atmosphere at 500$^\circ$C with the rates tuned to maximize the superconducting transition temperature. After the NbTiN growth, the substrate is cooled to room temperature without breaking vacuum and a 50\,nm layer of Al is deposited.  The wafer is then patterned using an i-line stepper. The Al is etched away using Transene A etchant, which does not attack the NbTiN beneath, only leaving Al in the bondpad regions. The wafer is then re-patterned and the NbTiN is etched using a CHF$_3$ based plasma in an ICP-RIE to define the center strip of the inverted microstrip. This etch does not quickly trench into the Si which could complicate later step coverage, and leaves a surface compatible with high $Q$ microwave devices~\cite{austermann2018-toltec}. An insulating layer of \mbox{$\alpha$-Si} is deposited at room temperature using ICP-PECVD, and then the ground plane of 100\,nm of Nb is sputtered.  The ground plane near the bondpads and transition from microstrip to CPW is patterned and etched using a CF$_4$-based RIE, which is selective versus the underlying \mbox{$\alpha$-Si}. Finally, the vias in the \mbox{$\alpha$-Si} are patterned and etched in an SF$_6$ RIE with the etch stopping on the Al covering the bond-pads. 

\section{Characterization Measurements}
On each wafer, we produce both KIT amplifier dies and diagnostic chips. The diagnostic chips are composed of lumped element resonator arrays to directly measure the NbTiN kinetic inductance and \mbox{$\alpha$-Si} permittivity. For the $L_k$ measurement we use an interdigitated capacitor and NbTiN straight segment to form the resonator, while for the permittivity measurement we use a parallel plate capacitor. The KIT dc bias is supplied to the devices using two bias-tees, while the pump signal is attenuated by 10\,dB at 4\,K and then delivered to the KIT package with a directional coupler. We measure a KIT critical current of 0.38\,mA and a scaling current of 2.1\,mA, which are compatible with values obtained in an earlier study with the same material~\cite{Giachero2023}. These values are lower compared to those measured for the $t=20$\,nm NbTiN CPW version~\cite{Malnou2021} ($I_c=2.4$\,mA and $I_*=7$\,mA) as expected.

\begin{figure}[!t] 
 \begin{center}
    \includegraphics[clip=true,width=\textwidth]{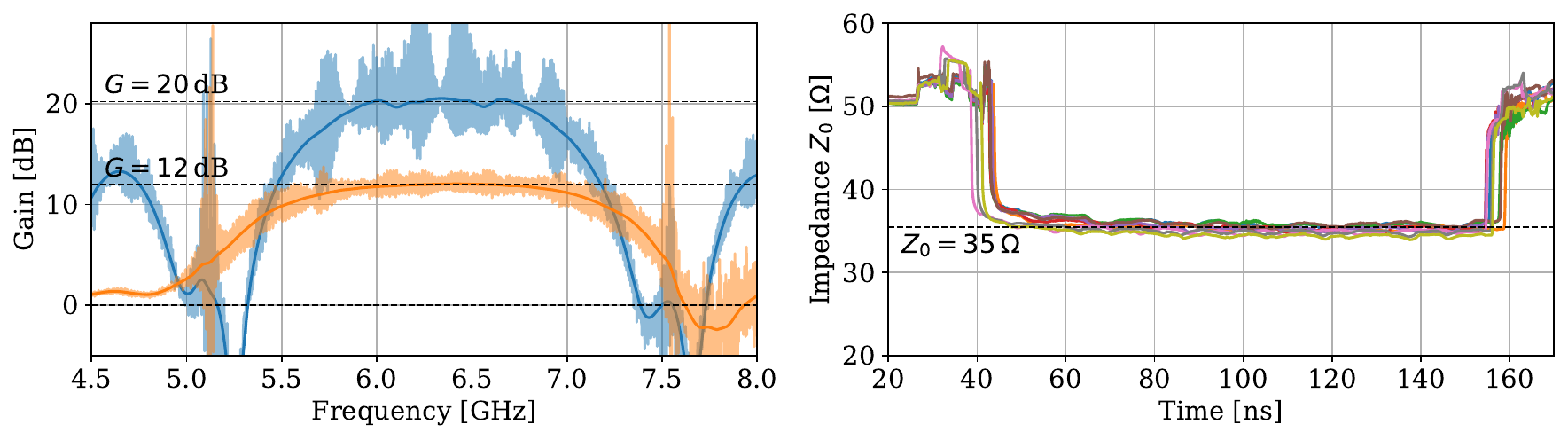}
  \end{center}
 \caption{\label{fig:gain_TDR} (left) Gain measured for two different configurations: 20\,dB gain has been obtained with $I_{dc}=0.24\,$mA, $f_p=12.662$\,GHz and KIT-input pump power around -31\,dBm, while 12\,dB gain has been obtained with $I_{dc}=0.13\,$mA, $f_p=12.666$\,GHz and -35\,dBm pump power. Note the 20\,dB gain settings result in significant levels of intermodulation products and higher order parametric processes, making operation in this regime less stable. Time-domain reflectometer measurement for eight amplifiers. The characteristic impedance is approximately $50\,\mathrm{\Omega}$ for all parts related to the characterization system, dropping down to $35\,\mathrm{\Omega}$ when the signal travels through the amplifiers. The impedance mismatch arises from a significantly higher $\varepsilon_r$ than was measured in an earlier process run and has been corrected for the next device revision.}
\end{figure}

Gain profile measurements are performed by determining the ratio between the forward transmission $S_{21}$ with fixed dc bias and the pump on and off. In total, eight devices were tested and all exhibited gain ranging between 10--20\,dB (Fig.~\ref{fig:gain_TDR}) and centered near 6\,GHz. During tune-up of the gain profile $I_{dc}$ ranged from 0.12 to 0.24\,mA, and the pump frequency is typically $f_p=12.6$\,GHz. Higher gain is possible with increased dc bias and pump power at the expense of a larger ripple. 

Typical pump powers are comparable to the previous 20\,nm NbTiN CPW version (-30\,dBm on-chip pump power)~\cite{Malnou2021}. However, the current devices have a much thinner NbTiN film so the required pump power is expected to be lower than the CPW devices and currently under investigation. We also discover an impedance mismatch of the microstrip line using a Time-domain Reflectometry (TDR) measurement which yields a characteristic impedance of $Z_0\sim 35\,\mathrm{\Omega}$ (Fig.~\ref{fig:gain_TDR}, right). With the diagnostic chips we measured the NbTiN film kinetic inductance to be $L_k = 30$\,pH/sq, and the \mbox{$\alpha$-Si} permittivity to be $\varepsilon = 9.6$. The latter is substantially discrepant with the expected value from past measurements. Simulations performed with the measured $L_k$ and $\varepsilon$, yielded a characteristic impedance around $Z_0 = 39\,\mathrm{\Omega}$.

\begin{figure}[!t] 
 \begin{center}
    \includegraphics[clip=true,width=0.80\textwidth]{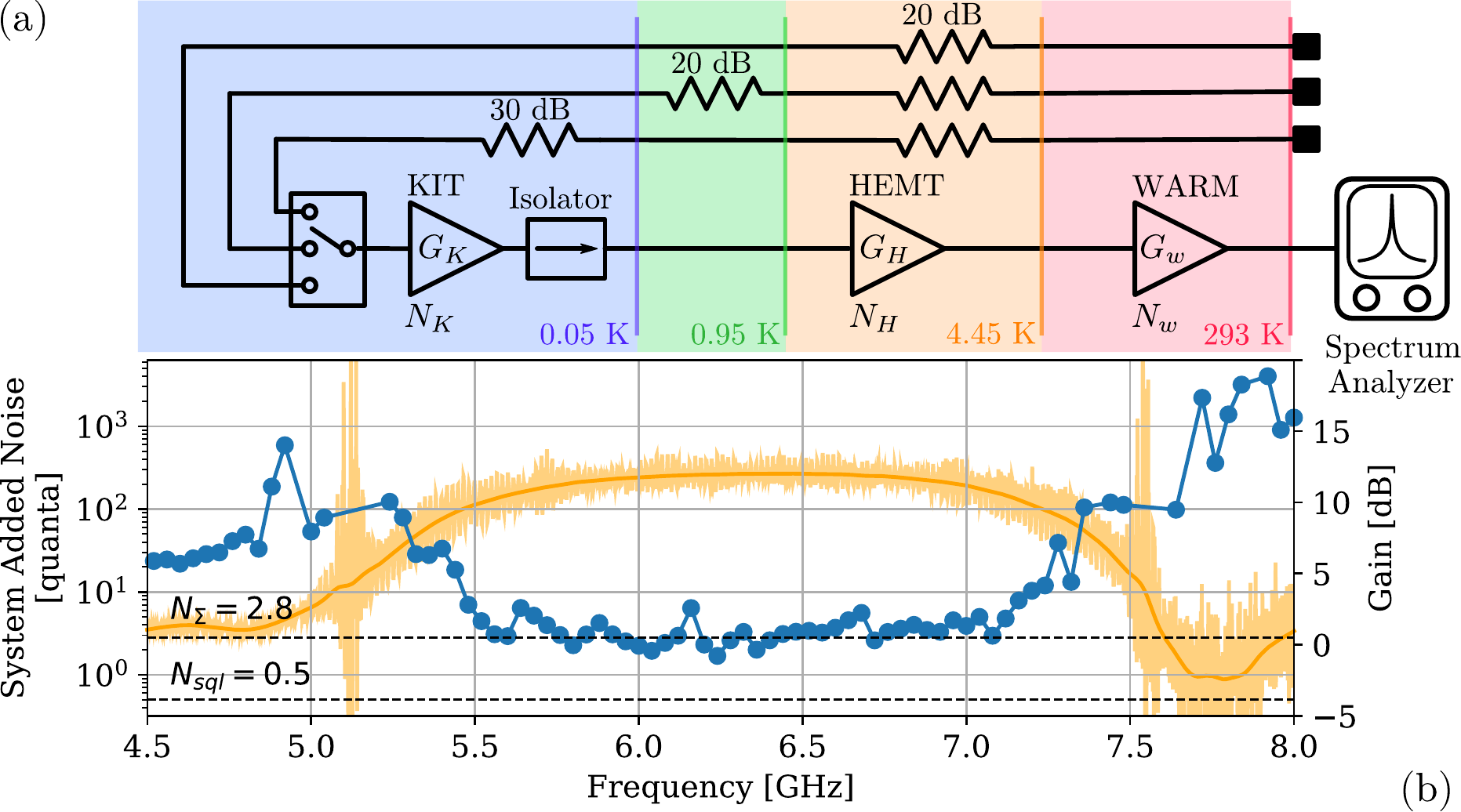}
  \end{center}
 \caption{\label{fig:noise_scheme_and_results}  A simplified dilution refrigerator schematic (a) used for the noise characterization. The KIT operates with gain $G_K$ and added noise $N_K$, and its output is amplified at 4.45\,K by a commercial HEMT amplifier with gain $G_K \sim 38$\,dB and added noise \mbox{$N_H = 3$--10\,quanta} (compatible with \cite{Malnou2021}) and again at room temperature with $G_W \sim 26$\,dB. We may neglect $N_W$ as the effective signal temperature at this stage is $\mathcal{O}(10^7)$\,K and the amplifier's noise figure is $\lesssim 6$. Use of a 20~dB isolator at 0.05~K, between the KIT output and HEMT input, allows us to neglect the HEMT's noise as an additional term in the calculation of the noise power at the KIT input for each switch position. (b) Estimated system-added noise (left vertical axis) and KIT gain (right vertical axis) as a function of the frequency. The darker yellow line shows the smoothed gain profile.}
\end{figure}

To characterize the system-added noise with the KIT as the first-stage amplifier we use a modified $y$-factor (``hot/cold") method using three coaxial lines with attenuators at three different temperature stages \{20, 20, 30\}~dB at \{4.45, 0.95, 0.05\}~K. These three noise temperatures are connected to the KIT via a cryogenic micro-electromechanical system (MEMS) switch. A simplified schematic circuit for the noise characterization is shown in Fig.~\ref{fig:noise_scheme_and_results}(a). Each noise input line is terminated at room temperature so consideration of the entire chain moving from 293~K to the KIT input is necessary to obtain the correct effective temperatures. After a full loss characterization of each component (all at base temperatures) we obtain mean input temperatures of \{3.41, 0.55, 0.16\}\,K, or \{11.0, 1.8, 0.7\} quanta across the
KIT bandwidth.

To estimate the KIT-on system-added noise we measure the system output power level using a Spectrum Analyzer (SA) configured in zero-span mode with 1\,MHz resolution and video bandwidths. We set the SA center frequency, gather ten traces, and actuate the MEMS switch between each of its three positions -- after which the center frequency is stepped and the procedure is repeated. At each SA frequency we perform a linear fit to the three-point curve corresponding to the system output noise level at each switch position. We then extract an estimate for the system-added noise, $N_\Sigma$ (units of quanta), via
\begin{equation}
    N_{out}^s = G_c(N_{in}^s + N_{in}^i + N_\Sigma).
\end{equation}
$G_c$ is the total chain gain and $N_{in}^{s, i}$ is the input noise quanta at the signal or idler frequencies. We use the high-gain approximation of $G_K \approx G_K - 1$ to simplify our analysis \cite{Malnou2021} and, leave $G_c$ as a free parameter in the fit. In this technique we neglect the HEMT-added noise, $N_H$ (measured concurrently to be \mbox{$N_H = 3$--10\,quanta} across the KIT bandwidth). Doing so effectively attributes the HEMT contribution to the total system-added noise as belonging to the KIT-added noise. In the case where $N_H$, when referred to the KIT input, is much less than the SQL (in our setup with 20\,dB isolation and a KIT gain of 13\,dB the HEMT noise at the KIT input is far below the SQL) this measurement yields an upper bound for the estimated system-added noise with the KIT as the first-stage amplifier. Fig~\ref{fig:noise_scheme_and_results}(b) shows the estimated system-added noise and demonstrates a 2.5~GHz 3\,dB bandwidth (\mbox{5.6--7.1\,GHz}) over which $N_\Sigma \leq 3$ quanta (860\,mK at 6\,GHz). As the experimental setup is quite similar to that of \cite{Malnou2021}, and we measure a comparable HEMT noise, these results indicate the new inverted microstrip KIT design operates with a similar proximity to the SQL as previous CPW devices.

\section{Conclusion and Future Plans}
We designed and produced a stub-loaded inverted microstrip Kinetic Inductance Traveling Wave Parametric Amplifier, utilizing a superconducting material with significantly higher kinetic inductance than prior devices. The increased kinetic inductance should lead to a reduction in the required pump power and dc bias for amplifier operation and will be carefully characterized in next-generation devices. Although the device's characteristic impedance is not well-matched to 50\,$\Omega$, we achieve gain over 10\,dB and a system-added noise around 3 quanta (860\,mK) at 6\,GHz. These promising results prompted the development of a new design optimized for a matched characteristic impedance, which has already been fabricated and is currently being characterized. Integration of the second generation KITs with rf-SQUID multiplexer readout devices used for TES detector readout, and assessment of the readout system performance is currently underway. Future steps towards improving these devices' practical use in detector readout chains involve implementation of on-chip bias-tees and directional couplers/diplexers. Such on-chip superconducting bias circuits minimize loss and reduce the total amplifier footprint; making them much easier to use in already complex detector systems. These compact, low-noise amplifiers, could be advantageous also for MKIDs, particularly in single-photon counting applications~\cite{Mezzena2020}, and for MMCs~\cite{Kempf2013}, which can be read out using rf-SQUID microwave multiplexing similar to those currently is use for TES readout.

\bmhead{Acknowledgments}
This work is supported by the National Aeronautics and Space Administration (NASA) under Grant No. NNH18ZDA001N-APRA, the Department of Energy (DOE) Accelerator and Detector Research Program under Grant No. 89243020SSC000058, and DARTWARS, a project funded by the European Union’s H2020-MSCA under Grant No. 101027746. The work is also supported by the Italian National Quantum Science and Technology Institute through the PNRR MUR Project under Grant PE0000023-NQSTI.

\bibliography{dartwars-LTD20_sn}

\end{document}